\documentclass[preprint,showpacs,preprintnumbers,amsmath,amssymb,nofootinbib]{revtex4}

\usepackage{etex}
\usepackage{graphicx}
\usepackage{dcolumn}
\usepackage{bm}
\usepackage{amssymb,amsthm,amscd,amsbsy,array}\usepackage{amsmath,dsfont}

\usepackage{graphics,xcolor}

\usepackage{color}
\newcommand{\red}{\textcolor{red}}  
\newcommand{\blue}{\textcolor{blue}}

\newcommand{\half}{{\scriptstyle{\frac{1}{2}}}}
\def\2{{\half}}
\newcommand{\PT}{{P\"oschl - Teller\;}}
\newcommand{\dPT}{{derived P\"oschl - Teller\;}}
\newcommand{\SL}{{Sturm-Liouville\;}}

\newcommand{\DM}{{Displacement Memory\;}}
\newcommand{\const}{\mathop{\rm const}\nolimits}
\def\parag{\hfil\break} 
\def\kikezd{\parag\underbar}
\def\bX{{\bm{X}}}

\def\p{{\partial}}
\def\bk{{\bf k}}

\def\bp{{\bm{p}}}
\def\bb{{\bf b}}

\def\bc{{\bf c}}

\def\bp{{\bm{p}}}

\def\bx{{\bm{x}}}
\def\bxi{{\bm{\xi}}}

\def\beq{\begin{equation}}
\def\eeq{\end{equation}}
\def\beqa{\begin{eqnarray}}
\def\eeqa{\end{eqnarray}}

\def\barray{\left(\begin{array}}
\def\earray{\end{array}\right)}
\def\barraynb{\begin{array}}
\def\earraynb{\end{array}}

\def\IS{{\mathfrak{S}}} 

\def\smallover#1/#2{\hbox{$\textstyle\frac{#1}{#2}$}}

\newcommand{\cA}{{\mathcal{A}}}

\newcommand{\Ker}{\mathrm{Ker}\,}

\def\smallcirc{{\raise 0.5pt \hbox{$\scriptstyle\circ$}}}
\def\aand{{\quad\text{\small and}\quad}}

\def\ie{{\;\text{\small i.e.}\;}}
\def\ie,{{\;\text{\small i.e.,}\;}}

\DeclareMathOperator{\sech}{sech}
\def\StL{{Sturm-Liouville\;}}

\def\benu{\begin{enumerate}}
\def\eenu{\end{enumerate}}
\def\bitem{\begin{itemize}}
\def\eitem{\end{itemize}}

\def\besub{\begin{subequations}}
\def\esub{\end{subequations}}

\def\?{{\,\gb{\fbox{\texttt{??}}\;}}\,}

\usepackage[colorlinks=true, pdfstartview=FitV, linkcolor=blue, citecolor=blue, urlcolor=blue]{hyperref} 
\newcommand{\brown}{\textcolor{brown}}
\newcommand{\cyan}{\textcolor{cyan}}
\newcommand{\magenta}{\textcolor{magenta}}
\newcommand{\violet}{\textcolor{violet}}

\newcommand{\gb}{\quad\colorbox{green}}

\newcommand{\dgreen}{\textcolor[rgb]{0,0.5,0}}

\newenvironment{redtext}{\color{red}}
{\ignorespacesafterend}
\newenvironment{bluetext}{\color{blue}}{\ignorespacesafterend}
\newenvironment{greentext}{\color{green}}{\ignorespacesafterend}
\newenvironment{magentatext}{\color{magenta}}{\ignorespacesafterend}
\newenvironment{cyantext}{\color{cyan}}{\ignorespacesafterend}
\newenvironment{orangetext}{\color{orange}}
{\ignorespacesafterend}

\newcommand{\bmagenta}{\begin{magentatext}}
\newcommand{\emagenta}{\end{magentatext}}
\newcommand{\bcyan}{\begin{cyantext}}
\newcommand{\ecyan}{\end{cyantext}}
\newcommand{\bblue}{\begin{bluetext}}
\newcommand{\eblue}{\end{bluetext}}
\newcommand{\bred}{\begin{redtext}}
\newcommand{\ered}{\end{redtext}}

\newcommand{\bgreen}{\begin{greentext}}
\newcommand{\egreen}{\end{greentext}}
\newcommand{\borange}{\begin{orangetext}}
\newcommand{\eorange}{\end{orangetext}}

\numberwithin{equation}{section}

\let\ssection=\section
\renewcommand{\section}{\setcounter{equation}{0}\ssection}

\newcommand{\bigbox}[1]{\fbox{%
\rule[-20pt]{0pt}{45pt}$\;\;\displaystyle{#1}\;\;$}
}
\newcommand{\medbox}[1]{\fbox{%
\rule[-10pt]{0pt}{25pt}$\;\;\displaystyle{#1}\;\;$}%
}

\usepackage{amsthm}

{Corollary}[section]

\begin{document}


\title{
Globally defined Carroll symmetry of gravitational waves 
}

\author{
M. Elbistan$^{1}$\footnote{mailto:mahmut.elbistan@bilgi.edu.tr},
P.-M. Zhang$^{1,2}$\footnote{mailto:zhangpm5@mail.sysu.edu.cn}
P. A. Horvathy$^{1,3}$\footnote{mailto:horvathy@univ-tours.fr},
}

\affiliation{
${}^1$ Department of Energy Systems Engineering, Istanbul Bilgi University, 34060, Eyupsultan, Istanbul, (Turkey)
\\
$^2$ School of Physics and Astronomy, Sun Yat-sen University, Zhuhai, China
\\
${}^3$ Institut Denis-Poisson CNRS/UMR 7013 - Universit\'e de Tours - Universit\'e d'Orl\'eans Parc de Grammont, 37200 Tours, (France).
\\
}

\date{\today}

\begin{abstract}  
The local Carroll symmetry of a gravitational wave found in Baldwin-Jeffery-Rosen coordinates is extended to a globally defined one by switching to Brinkmann coordinates. Two independent globally defined solutions of a Sturm-Liouville equation allow us to describe both the symmetries (translations and Carroll boosts) and the geodesic motions. One of them satisfies particular initial conditions which imply zero initial momentum, while the other does not. Pure displacement arises when the latter is turned off by requiring the momentum to vanish and when the wave parameters take, in addition, some particular values  which correspond to having an integer half-wave number. The relation to the Schwarzian derivative is highlighted. We illustrate our general statements by the P\"oschl-Teller profile.
\\

Nuclear Physics B (accepted).
\end{abstract}


\maketitle

\tableofcontents

\section{Introduction}\label{Prelim}

Braginsky and Thorne \cite{BraTho} argued that gravitational waves might be observed by detecting the displacement of particles initially in rest, called later the \emph{Memory Effect}.  
While experimental verification is still in the making, conceptual insight can be gained by taking advantage of the \emph{Carroll symmetry} \cite{LeblondCar,SenGupta}. 
Baldwin-Jeffery-Rosen  
(BJR) coordinates \cite{BaJe,Rosen}  provide a simple description \cite{Sou73,Carroll4GW}, however are regular only in finite intervals, requiring to glue together the local  results. 

In this paper we show that such complications can be avoided by switching to globally defined Brinkmann (B) coordinates \cite{Bri,DGH91}. The transcription requires to solve a matrix \SL equation,  \eqref{SL+cond} below. 
 The price to pay for globality is to get more complicated-looking expressions, though. 
The advantages and drawbacks of the two coordinate systems is discussed in the Conclusion.

We illustrate our general theory by the P\"oschl-Teller profile  \cite{PTeller,Plyushchay09,ChaKar}, which is a good \emph{analytical} approximation of the widely studied Gaussian one.
The book \cite{exactsol} and our review \cite{EZHRev}  provide the reader with further references.

Our notations are: $(\bX,U,V)$ are Brinkmann  coordinates and $(\bx,u,v)$ are BJR coordinates. $U=u$ is an affine parameter for geodesics. $\Theta^{Brink}=\bxi\p_{\bX}+\eta\p_V$ denotes collectively the vector fields which generate the infinitesimal isometries of spacetime. Its explicit form will be spelt out in \eqref{CarrollBrink}.

\section{From Brinkmann to BJR and back}
\label{BriBJR}

 Plane gravitational waves are conveniently described by using globally defined Brinkmann coordinates \cite{Bri,DGH91,exactsol} in terms of which the metric is, 
\beq
g_{\mu\nu} dX^{\mu} dX^{\nu} =d\bX^2+2dUdV+K_{ij}X^iX^j%
\,dU^2\,,  
\label{Bplanewave}
\eeq
where the 
 $\bX = (X^1,\dots,X^D)$ are coordinates of the  transverse plane with the flat Euclidean metric $d\bX^2=\delta_{ij}\,dX^idX^j$. $U$ and $V$ are light-cone coordinates related to the  usual relativistic coordinates according to  
 $U = \frac{1}{\sqrt{2}} (Z-T)$ and $V = \frac{1}{\sqrt{2}} (Z+T)$. The profile
 $(K_{ij})$ is a symmetric $D\times D$ matrix whose entries depend only on $U$. Henceforth we focus our attention at $D=1$ or $D=2$ and assume, for simplicity, that the wave is linearly polarized so that $(K_{ij})$ is diagonal. Thus in $D=2$, which is our main interest in this paper, 
\beq
K_{ij}X^iX^j=\cA(U) \Big((X^{+})^2-(X^{-})^2\Big)\,,
\eeq
 which yields a vacuum solution of the Einstein equations for any $\cA(U)$ \cite{Bri,exactsol,DGH91,EZHRev}. We consider only ``sandwich waves'' \cite{BoPi89}, whose profile $\cA(U)$ vanishes outside an interval $U_i< U < U_f$. 
Choosing $U$ as an affine parameter, the equations of motion of the transverse coordinates are of the \StL form,
\beq
\dfrac{d^2\bX}{dU^2}-\barray{lr}
\cA &0
\\
0  
& -\cA \earray
\,\bX = 0\,.
\label{ABXeq}
\eeq 

The geodesic equations  have an additional constant of the motion called the Jacobi invariant \cite{exactsol},
\beq
e=\frac{1}{2}\,g_{\mu\nu}\,\frac{d{X}^\mu}{dU}\frac{d{X}^\nu}{dU}\,.
\label{Jacobi}
\eeq
The transverse equations \eqref{ABXeq} do not depend on $e$.
In this paper we limit our attention at the lightlike case $e=0$ for which the  ``vertical'' coordinate $V$ satisfies 
\beq
\dfrac{d^2V}{dU^2}+ \dfrac{1}{2}
\displaystyle{\frac{d\cA}{dU\,}}\Big((X^1)^2-(X^2)^2\Big)
+2\cA\Big(X^1\dfrac{dX^1}{dU\,}-X^2\dfrac{dX^2}{dU\,}\Big)=0\,.
\label{ABVeq}
\eeq
It corresponds to the horizontal lift of  \eqref{ABXeq}
and can be solved once the solutions of the latter are known.
Henceforth we focus our attention at \eqref{ABXeq}.  The timelike  (massive) case, $e < 0$, is discussed, e.g. in \cite{EDAHK,DM-1}.
The equations above constitute a model for particle dynamics in a gravitational wave including flyby \cite{ZelPol}.
  
After the  wave has passed,  particles initially at rest exhibit  the \emph{Velocity Memory Effect} (VM) \cite{Ehlers,AiBalasin, EZHRev}~: outside the Wavezone, they move with constant velocity. Under certain ``quantization'' conditions the outgoing velocity can vanish, though, and we get  the \emph{Displacement Memory Effect} (DM)  \cite{ZelPol,DM-1,DM-2,Sila-PLB}. 

Plane gravitational waves have long been known to have a $2D+1$ parameter symmetry group composed of $D+1$ translations, completed by $D$ (rather mysterious) transformations \cite{BoPiRo,Sou73}. More recently  \cite{Carroll4GW}, this group was identified with  the \emph{Carroll group} \cite{LeblondCar,SenGupta}, as young L\'evy-Leblond called it jokingly. 
 Its algebraic structure is readily determined by switching to Baldwin-Jeffery-Rosen (BJR) coordinates \cite{BaJe,Rosen,Sou73} which are however only local.  
This note sheds further light at the Carroll --- Memory relation using globally defined Brinkmann coordinates.

\goodbreak

We start with considering real solutions of the Sturm-Liouville equations
\beq
{P}''= \cA(U)P,
\qquad
 (P^T){\,}{P}'=(P^T)'{\,}P\,
\label{SL+cond}
\eeq
for a $D{\times}D$  matrix $P(U)=\big(P_{ij}(U)\big)$  \cite{EZHRev}. The prime here denotes $d/du \equiv d/dU$.
BJR coordinates $(\bx, u,v)$ are then obtained by
\beq
{\bX} = P(u)\bx,
\qquad
U=u,
\qquad 
V=v-\frac{1}{4}\bx\cdot \big(P^T{}P\big)'\bx\,,
\label{BBJRcoord}
\eeq
in terms of which the metric takes the BJR form,
\beq
g_{\mu\nu}dx^{\mu}dx^{\nu}= \left(P(u)^TP(u)\right)_{ij}(u)\,dx^idx^j+2du\,dv\,.
\label{nopotform}
\eeq

Carroll symmetry written in BJR coordinates is composed of ``vertical'' ($v$) and transverse ($\bx$)  translations with parameters  $h$ and $\bc$, respectively, completed by Carroll boosts with parameter $\bb$ \cite{Sou73, Carroll4GW}, 
\beq
\bx\to\bx+\bc+ S(u)\,\bb\,,
\qquad
u\to u,
\qquad
v\to v-\bb\cdot\bx - \2\bb\cdot{}S(u)\,\bb+h\,.
\label{BJRCarr}
\eeq
The symmetric $2\times 2$ matrix
\beq
S(u)=\int^u_{u_0}\!\left(P^TP\right)^{-1}\!(t) dt
\label{Smatrix}
\eeq
here is referred to as the \emph{Souriau matrix} \cite{Sou73,Carroll4GW,EZHRev}. $S$ is related to but is slightly different from the time redefinition 
 considered, in the \emph{isotropic case}, by Arnold \cite{Arnold,ZurabNote,Aniso,ZZH}. 
 Details will be as discussed elsewhere \cite{ZEHPR}.
 
The infinitesimal symmetries are,  with some abuse of notation,
\begin{equation}
\label{CarBJR}
\theta^{BJR} = h\partial_v + c^j\partial_j + b_j \big(S^{ji}\partial_i - x^j\partial_v\big)\,.
\end{equation}
The covariantly constant Killing vector $\p_v$  here generates ``vertical'' translations -- a hallmark of Brinkmann metrics \cite{Bri,DGH91}. The $c_i$ generate translations and the $b_j$ generate  Carroll boosts. For $u=u_0$, the integral \eqref{Smatrix}  vanishes and we recover the usual Carroll action in \cite{LeblondCar}.
 
Finding the geodesic trajectories is then straightforward by using the symmetry. 
Noether's theorem provides us with  associated conserved quantities \cite{Sou73,Carroll4GW},
\beq
\bp=\bp_0= (P^TP)(u)\,{\bx}'(u),
\qquad
\bk=\bk_0 = \bx(u)-S(u)\,\bp\,,
\label{CarCons}
\eeq
 interpreted as conserved \emph{linear and  boost-momentum},
 respectively, supplemented by the non-relativistic mass that we scale to $1$. It is generated by the Killing vector $\p_v$. 
The conserved quantities determine the geodesic trajctories \footnote{A non-vanishing Jacobi invariant $e\neq0$ would merely add a linear term,
$e u$ to the r.h.s., \cite{EDAHK,DM-1}.},
\beq
\bx(u) = \bk_0 + S(u)\,\bp_0\,,
\quad v(u) = -\frac{1}{2} \bx.\bp +  v_0\,. 
\label{BJRtraj}
\eeq
  
So far so good. However a subtlety comes from that the BJR coordinates are defined only in intervals $I_k = [u_{k-1},u_k]$ between  \emph{mandatory} \cite{Sou73,Carroll4GW} zeros of the determinant\footnote{In $D=1$, $P(U)$ is just a scalar function.}, 
\beq
\det P(u_k)  = 0\,.
\label{detP0}
\eeq
The matrix $P$ plays a fundamentally important r\^ole in our investigation, as we shall see. Both the metric \eqref{nopotform} and the Souriau matrix $S$  \eqref{Smatrix} are singular at the junction points $u_k$, as is manifest 
 in FIG.\ref{BPSfig} for the \PT profile.
 The singularity  will be removed  by switching to Brinkmann coordinates.

Although our results could be extended to non-diagonal profiles and to any dimension, 
henceforth we focus our attention at the diagonal case, $P^T = P$, and  to  $D=2$ \ie, to the physically relevant $4$-dimensional spacetime. Occasionally, we consider also toy examples in $D=1$, though.
\goodbreak
   
Let us assume that $u < u_1$, the first zero of $\det(P)$. 
Requiring that our particles be in rest before the wave arrives, 
\beq
{\bx}(-\infty)=x_0  \aand {\bx}'(-\infty)=0\,,
\label{xinit}
\eeq
implies that the momentum vanishes  in $I_1$, 
 \beq
\bp=0\,.
\label{0mom}
\eeq
The Souriau term is thus switched off from \eqref{BJRtraj} and the transverse BJR ``trajectory'' is, for vanishing Jacobi invariant, $e=0$ in \eqref{Jacobi}, merely a fixed point \footnote{A non-vanishing Jacobi in $e\neq0$ would merely shift $v(u)$ by a linear-in-$u$ term \cite{EDAHK,DM-1}.}, 
\beq
\bx(u)=\bx_0=\const \,, \qquad
v(u)=v_0=\const\,,
\label{BJRnomotion}
\eeq
consistently with \eqref{BJRtraj}. 
Eqn \eqref{BJRnomotion} confirms the  ``no  motion for Carroll" wisdom \cite{LeblondCar}~\footnote{Motion is possible for multi-particle systems \cite{MultiCarroll}, and for planar models with ``exotic'' double central extension when the particles follow the Hall law \cite{MARSOT,ExoPeierls}.}. 
However, unlike as for BJR,  the (lightlike) Brinkmann trajectory is  \emph{non-trivial}~: \eqref{BBJRcoord} yields, 
\beq
\bX(U) = P(u)\,\bx_0\,,
\qquad
V(U) = -\frac{1}{2}\bX.\bX' + V_0\,, 
\label{BDMtraj}
\eeq
with $U=u$ and $V_0=v_0$. The initial conditions \eqref{xinit} then require :
\beq
P(-\infty)= {\rm Id} \aand P'(-\infty)=0\, .
\label{Pinit}
\eeq

The Brinkmann trajectory, given by the action of the $P$-matrix, \eqref{BDMtraj}, may look indeed quite complicated, depending on the \SL solution $P$ of \eqref{SL+cond}  \cite{EZHRev}.
For randomly chosen wave parameters we get VM
 \cite{Ehlers,AiBalasin,EZHRev}~: the particles fly apart with constant velocity. 
For appropriately ``quantized'' values of the wave parameter, though,  
the wave zone contains an integer number of half-waves;  
 $P(u)$ then tends to a constant matrix also for $u=U\to\infty$ and DM is obtained \cite{DM-1,DM-2,Sila-PLB}. 
 
We emphasise, though, that the investigations above are {\rm a priori} valid only before the first zero of $\det{P}$ and should then be restarted until the next zero, and so on, as it will be seen in sec.\ref{PTsec} for \PT.

Another remarkable consequence of \eqref{BDMtraj} 
 is that when $\det P(u_0) = 0$ for some $u_0$, then \emph{all} transverse trajectories  which start from  $x_0 \in \Ker \big(P(u_0)\big) $ are focused at the origin~: we get a \emph{caustic point} \cite{BoPi89,EZHRev}, as it will be illustrated in  FIG. \ref{Btrajm1m2}.
 
\section{Carroll symmetry in Brinkmann coordinates}\label{BrinkmannCarrollSec}

Simple as they are, BJR coordinates  \eqref{BBJRcoord} suffer from being valid only in intervals  
between subsequent zeros of $det(P)$. The adjacent expressions should be glued together -- which may be laborious. A global approach can conveniently be given instead by using Brinkmann coordinates.
For simplicity we restrict our attention at $D=1$ transverse dimension and start with a fixed interval $I_k$. 
A straightforward computation then shows that pulling back the BJR expression \eqref{BJRCarr} by   \eqref{BBJRcoord} [or its infinitesimal form \eqref{CarBJR}] yields the Carroll generators written in Brinkmann coordinates, 
\begin{equation}
\bigbox{
\Theta^{Brink} = h \frac{\partial}{\partial V} + c \left(P \frac{\partial}{\partial X} - P' X \frac{\partial}{\partial V}\right) + b \left((PS)  \frac{\partial}{\partial X} - (PS)' X  \frac{\partial}{\partial V}\right)\,,
}
\label{CarrollBrink}
\end{equation}
which could also be verified independently, by checking the  equation $L_Yg = 0$ \cite{Torre, GenRelGrav}. 
The coefficients of $\p_X$ and of $\p_V$ are indeed the  symmetry generators $\xi$ and $\eta$ mentioned in the Introduction. 
They will be spelt out explicitly for \PT in Sec.\ref{PTsec}.
 The $P$-matrix and its derivative $P'$ generate translations, whereas $Q=PS$ and  $Q'=(PS)'$  generate Carroll boosts \cite{GenRelGrav}.  $\p_V$ generates the non-relativistic mass  scaled to $1$, as does for BJR.
 
 The only nonzero commutator 
yields a vertical translation \cite{Blau24},
\begin{equation}
\Big[\underbrace{P \frac{\partial}{\partial X} - P' X \frac{\partial}{\partial V}}_{translation}\ , \  
\underbrace{PS  \frac{\partial}{\partial X} - (PS)' X  \frac{\partial}{\partial V}}_{boost}\Big] = -\partial_V\,. 
\label{KillComm}
\end{equation}
The symmetry is thus the Heisenberg algebra.

$P(U)$ thus plays a double role~: for DM parameters it  determines the complicated-looking Brinkmann trajectory in \eqref{BDMtraj}, whereas by \eqref{CarrollBrink}  it also generates translations. 
For non-DM parameters, the situation is even more complicated, as will be seen  in \eqref{XPQ}. 

The Brinkmann form \eqref{CarrollBrink} has an important advantage w.r.t. the BJR expression  \eqref{BJRCarr}~: left multiplication by $P$ \emph{removes the singularity} of the Souriau matrix $S$ \cite{Sou73,EZHRev}~: 
\beq
Q(U) = P(U)S(U)\,
\label{2ndSLsol}
\eeq
is regular \emph{for all $U$}.  
In conclusion, switching form BJR to Brinkmann, \emph{\eqref{CarrollBrink},extends the  Carroll symmetry \eqref{BJRCarr} from an interval $I_k$ to the entire $U$-axis}.
A remarkable observation says that \emph{both $P$ and $Q$} are solutions of the  \SL equation \eqref{SL+cond}, 
\beq
{Q}''= \cA(U)Q,
\qquad
 (Q^T){\,}{Q}'=(Q^T)'{\,}Q\,,
\label{QSL+cond}
\eeq
as it can also be verified by a direct calculation.
$P$ and $Q$ have different initial conditions, though \cite{EZHRev,Jibril,Blau24}.     
The two solutions are independent when their Wronskian, $W(P,Q)= P' Q - Q' P$, does not vanish. 
\goodbreak
 
Contracting  the Brinkmann metric \eqref{Bplanewave} with the Carroll generators \eqref{CarrollBrink}  provides us  with,
\beq
\bp_0= P\bX'-\big(P\big)' \bX
 \aand
\bk_0= -Q\bX' + \big(Q\big)'\bX\,,
\label{Cconsquant}
\eeq
where $\bp_0$ and $\bk_0$ are, {\rm a priori}, new constants of the motion. However expressing them 
in BJR terms using \eqref{BBJRcoord}, the previous expressions  \eqref{CarCons} are recovered, as anticipated by our notation. 

The \SL solutions $P$ and $Q$ play yet another r\^ole, though~: either
combining \eqref{BJRtraj} with \eqref{BBJRcoord} or solving \eqref{Cconsquant} directly yields the general Brinkmann trajectory,
\beq\medbox{
\bX(U) = P(U)\, \bk_0 + Q(U) \,\bp_0 \,. 
}
\label{XPQ}
\eeq 

Remember that  $P$  satisfies the initial conditions \eqref{Pinit}. 

For vanishing momentum $\bp_0=0$, cf. \eqref{0mom},  the VM term $Q$ is switched off and  we recover \eqref{BDMtraj} with $\bx_0=\bk_0$. 
The corresponding Brinkmann trajectory may or may not be DM yet, though, depending on the wave profile. For parameters for which we have an integer number of half-waves in the wavezone, we  do get DM~: $P(+\infty)=\const$, though \cite{DM-1,DM-2,Sila-PLB}.
The case of nonzero initial momentum, $\bp_0\neq0$, will be studied in \cite{ZEHPR}.

\section{Schwarzian aspects}\label{SchwarzianSec}

For simplicity, we work in $D=1$.
Our results have an interesting relation with the 
 \emph{Schwarzian derivative} defined for a function $f$ 
as \cite{Ovsienko,Vlahakis,ZZH},
\beq
\IS\big(f(x)\big)= \frac{f'''(x)}{f'(x)}-\frac{3}{2}
\left(\frac{f''(x)}{f'(x)}\right)^2\,.
\label{Schwarziander}
\eeq

We note for later use that the Schwarzian depends only on  $f'$, not on $f$ itself.

Consider now the \SL equation
\beq
\varphi''(x)-\cA(x)\varphi=0\,,
\label{phiSL}
\eeq
cf.  \eqref{SL+cond} for both $P$  or for $Q$ collectively. Let then $\varphi_1(x)$ and $\varphi_2(x)$
be two independent solutions and consider their quotient,
\beq
 f(x)= \frac{\varphi_1(x)}{\varphi_2(x)}\,.
\label{phiperphi}
\eeq
It is then shown in \cite{Ovsienko} that the \SL profile can be recovered from the solutions in terms of the Schwarzian derivative,
\beq
\medbox{
\cA = - \half \,\IS(f)\,.
}
\label{Schwarzprofile}
\eeq

Let us now apply these general formulas 
 with the cast $ x \leadsto U$, completed by,
\beq
\varphi_1(U) = P(U)
 \aand 
\varphi_2(U) = P(U)S(U)\,.
\label{phiP}
\eeq
Thus 
\beq
f=\frac{1}{S}\,.
\label{fperS}
\eeq
The Wronskian $W(P,Q)$ is a constant, implying that
$f'$ in 
 \eqref{Schwarziander} is indeed proportional to  
$ S^{-2}.$ 
Thus by \eqref{Schwarzprofile} the profile is the Schwarzian derivative of the inverse of the Souriau matrix,
\beq
\cA = -\frac{1}{2}\, \IS\left(\frac{1}{S}\right)\,. 
\label{proffromS}
\eeq
The form \eqref{Smatrix} of $S$ then implies that  $P$  determines also the second solution we called $Q$.

Remarkably, we can also interchange the cast:
choosing 
\beq
\varphi_1(U) = Q=P(U)S(U)
 \aand 
\varphi_2(U) = P(U)\,
\label{phiQ}
\eeq
instead of \eqref{phiP},
it is $P$ which drops out, leaving us with the reciprocal relation for
\beq
\tilde{f} = \frac{Q}{P} = S\,,
\label{fQP}
\eeq
instead of $f= P/Q =1/S$ as  in \eqref{fperS}. 

The  new cast \eqref{phiQ} then yields, once again, a SL eqn. \eqref{phiSL}
whose profile is determined by the Souriau matrix,
\begin{equation}
\label{SouriauSch}
\cA(U) = -2\IS(S) 
= -2
 \frac{S'''(U)}{S'(U)} + {3}\left(\frac{S''(U)}{S'^(U)}\right)^2\,.
\end{equation}
which can indeed be viewed as an equation for $S$.
Somewhat surprisingly, 
inserting here $S$ from \eqref{Smatrix}, the  previous \SL equation \eqref{SL+cond} with the \emph{same} profile $\cA$ is recovered.
These results indicate that the interchange
\beq
P, \, Q \to   Q, \, P \;\;\;\Rightarrow\;\; \;
S \to \frac{1}{S}
\label{PQinterchnge}
\eeq
is a symmetry of the system, as it follows from the reciprocity relation
\beq
\IS(f) =  \IS\left(\frac{1}{f}\right)
\label{recip}
\eeq 
of the Schwarzian.
The difference between the two casts comes from the 
initial conditions. Those for $P$ are dictated by the 
behaviour at $U=-\infty$ but those for $Q$ involve also $S$ in \eqref{Smatrix}. 
   
\section{Illustration by the \PT profile}\label{PTsec}

\begin{figure}[h]
\includegraphics[scale=.33]{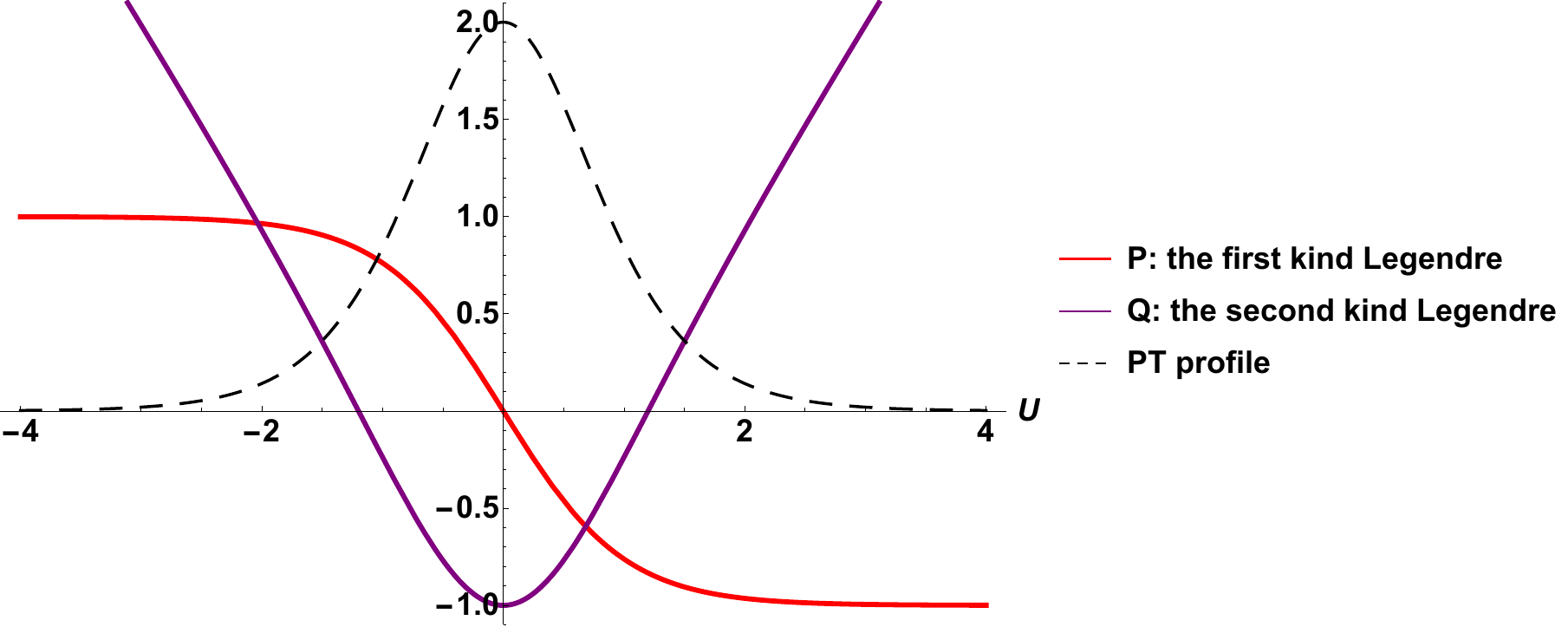}%
\vskip-3mm
\caption{
\textit{\small  
The solutions of the \StL eqn. \eqref{SL+cond} for  \PT\!,  shown for DM parameters. With initial conditions \eqref{Pinit} we get Legendre polynomials, \red{$P(U)=P_m(\tanh U)$}.  
 The second, independent solution, \violet{$Q(U)=Q_m(\tanh U)$}, is a Legendre function which does not satisfy the initial condition \eqref{Pinit}.} 
\label{PandPS} 
}
\end{figure}

We illustrate our general theory  by the \PT  profile \cite{PTeller,Plyushchay09,ChaKar,DM-1} in $D=1$,
\beq
\cA \equiv \cA^{PT}(U) =  -\dfrac{m(m+1)}{\cosh^2 U}\,.
\label{PTPot}
\eeq
The \SL equation \eqref{SL+cond} is solved by Legendre functions.  
 DM trajectories are obtained when  $m$ is an integer and we get Legendre \emph{polynomials}, 
 \beq
P(U) = (-1)^mP_m(\tanh U)\,.
 \label{DM4PT}
 \eeq 
The general geodesic,  \eqref{XPQ}, 
 combines the  DM solution $P(U)$ and the  non-polynomial $Q(U)$ in \eqref{2ndSLsol}, shown FIG.\ref{PandPS}. 

\begin{figure}[ht]
\includegraphics[scale=.35]{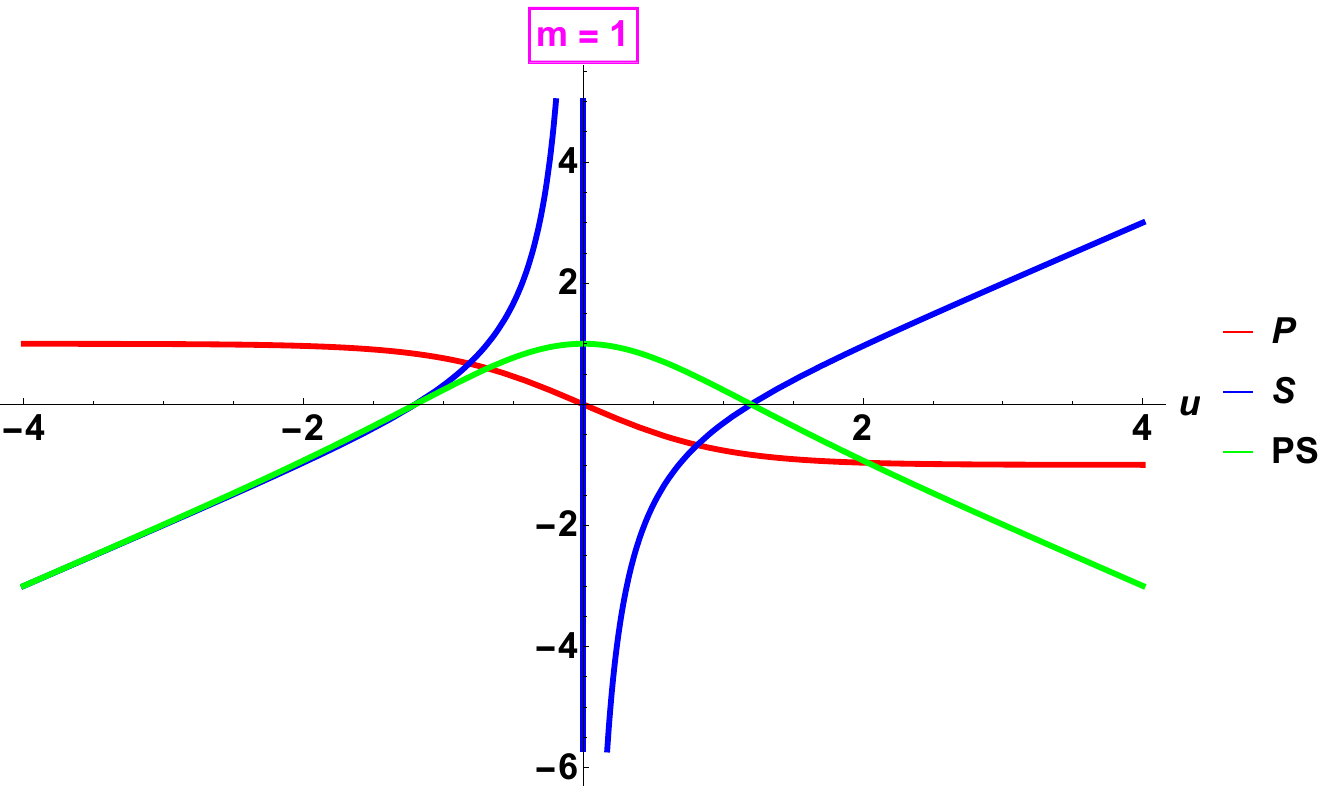}\;
\includegraphics[scale=.35]{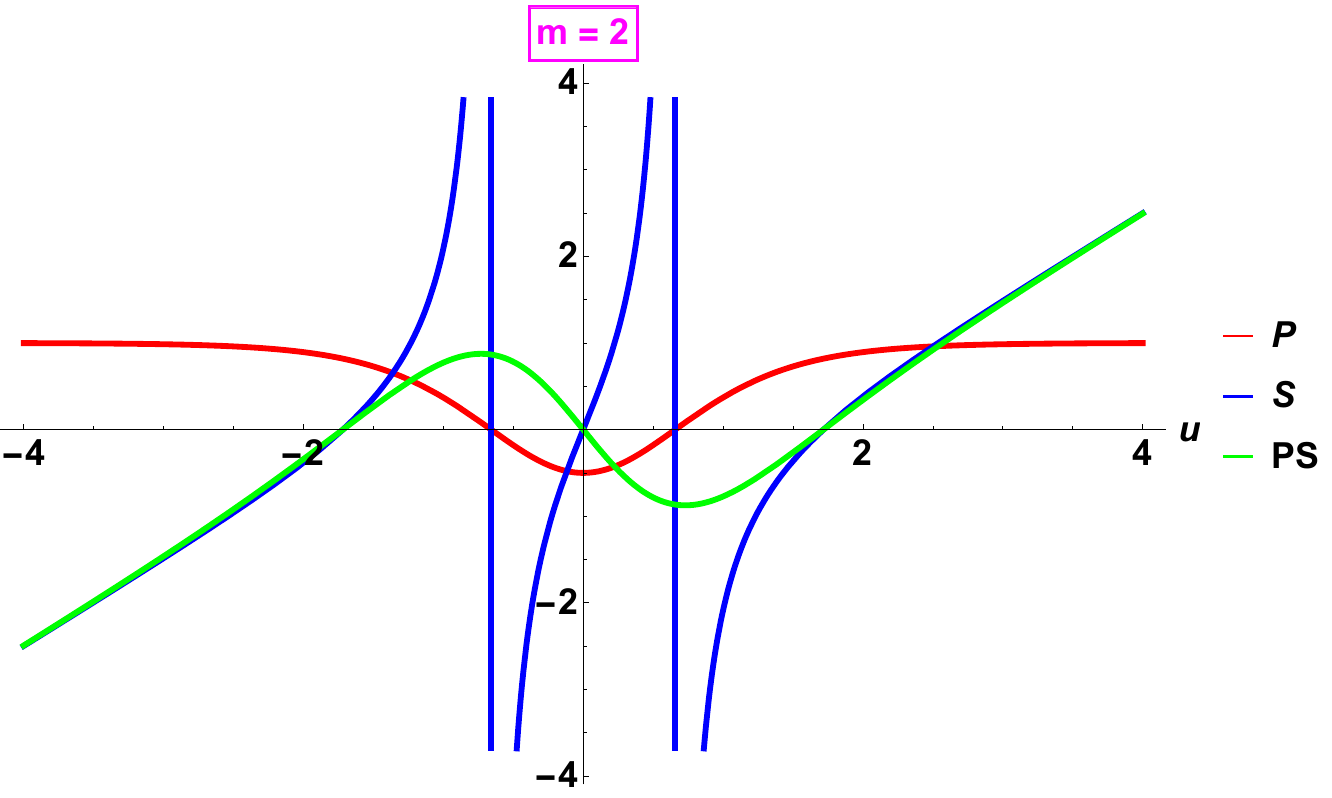}
\\
\vskip-5mm \caption{\textit{\small  
The  \SL solutions    \red{$\bf P$} of \eqref{SL+cond} shown   with wave numbers ${\bf m=1}$ and ${\bf m=2}$ for \PT in $D=1$. They yield  also DM trajectories written in Brinkmann coordinates.  
The Souriau matrix \blue{\bf S} diverges where \red{$\bf P=0$}, but  the pull-back to Brinkmann $\dgreen{\bf Q = PS}$ is regular for all $U$. 
}
\label{BPSfig}
}
\end{figure}

\medskip 
Turning to BJR cf. \eqref{BBJRcoord}, the Souriau matrix $S$ in \eqref{Smatrix} is regular between two subsequent zeros of $P$ but diverges at the junction points. For $m=1$, for example, 
\beq
S_{m=1}(u)=u-\coth u\,
\label{Smatrix1}
\eeq
obtained for the choice $u^{\pm}_{0}\approx \pm 1.2$,  which is  regular both in $I_{-} = \big\{u < 0\big)\}$ and in $I_{+} = \big\{u > 0\big)\}$, but diverges at $u=0$, as depicted in FIG.\ref{BPSfig}a. 
Similarly for $m=2$ shown in FIG.\ref{BPSfig}b,
\beq
S_{m=2}=\frac{1}{4}\left(u+\frac{3\sinh \left(2u\right)}{2(2-\cosh(2u))}\right)\, 
\label{Smatrix2}
\eeq
is singular where the denominator vanishes.

\begin{figure}[h]
\includegraphics[scale=.3]{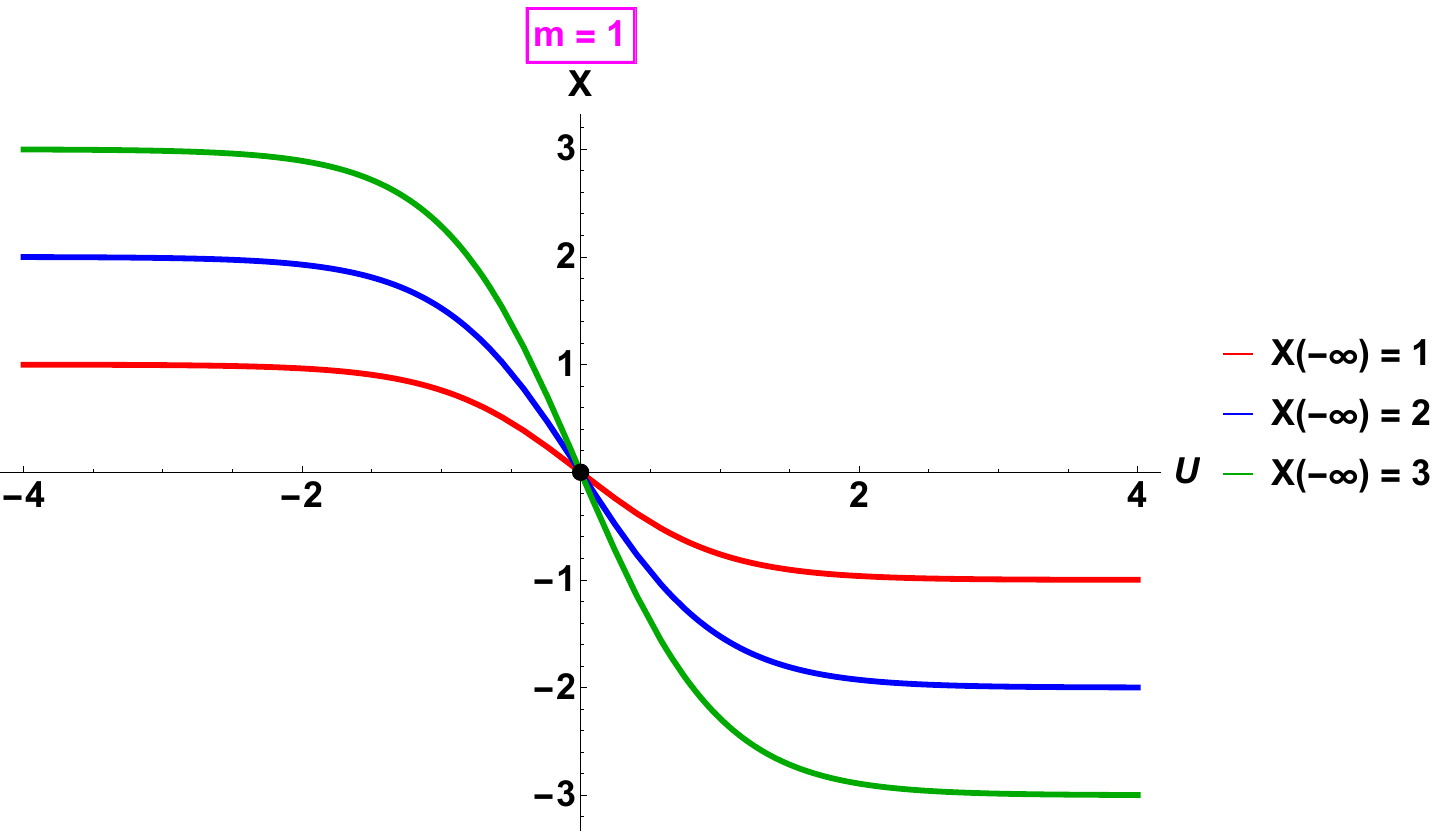}\quad
\includegraphics[scale=.3]{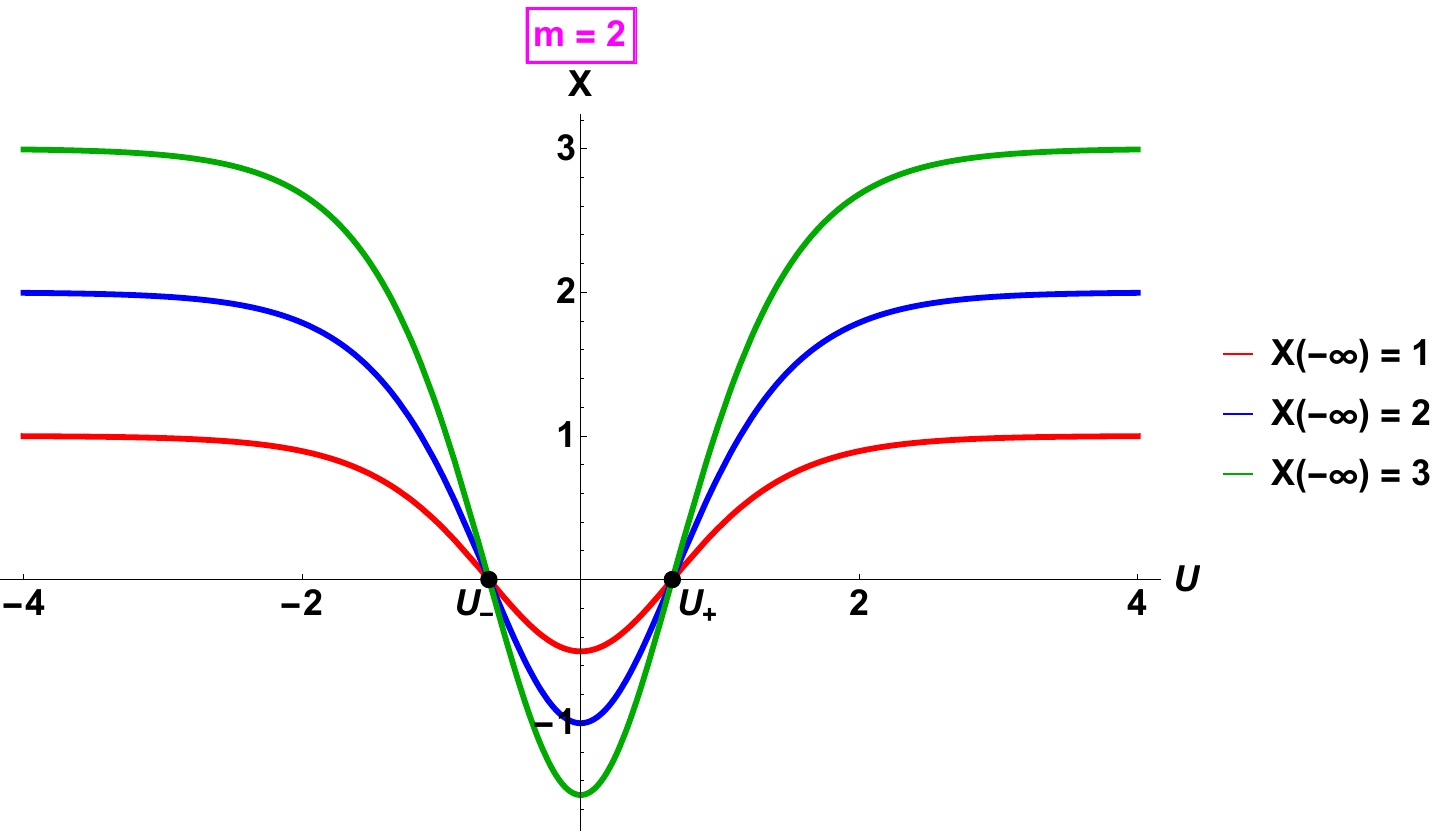}\vskip-3mm\caption{\textit{\small For a DM wave in $D=1$ transverse dimension  Brinkmann trajectories focus to the caustic points determined by $ P_m(U_k)=0$, $k=1,\dots,m$. For $m=1$ there is one focal point, at the origin, and for $m=2$, we have two of them, determined by $\tanh U_{\pm}=\pm{1}/{\sqrt{3}}$.} 
\label{Btrajm1m2} }
\end{figure}

$Q_m=P_mS_m$ in \eqref{2ndSLsol}, shown in FIG.\ref{BPSfig} in \dgreen{\bf green}, is a second solution of the \SL equation \eqref{SL+cond}. It is regular for all $U$, vanishes where the Souriau matrix does, and exhibits VM behaviour outside the Wave zone. 
$P_m$ and $Q_m$ are independent when the Wronskian does not vanish, $W(P,Q)\neq0$.
For $m=1$ and $2$ we have, for example,
\beq
\barraynb{llclllc}
\; P_1(U) &=& {-}\tanh U\,,   &\qquad 
&P_2(U) &=&\frac{1}{2}(3\tanh^2 U-1)\,, 
\\[6pt]
Q_1(U) &=&  {1-U\tanh U}, &\qquad
&Q_2(U) &=&\frac{1}{2}\big(2U - 3\tanh U - 3U\sech^2U\big)\,,
\label{BPQm1m2}
\earraynb
\eeq  
The $P(U)$ found above provide us with DM geodesics which are consistent with  \eqref{BDMtraj}. The zeros of $P$ are caustic points as it is manifest in FIG.\ref{Btrajm1m2}.

\begin{figure}[h]
\includegraphics[scale=.3]{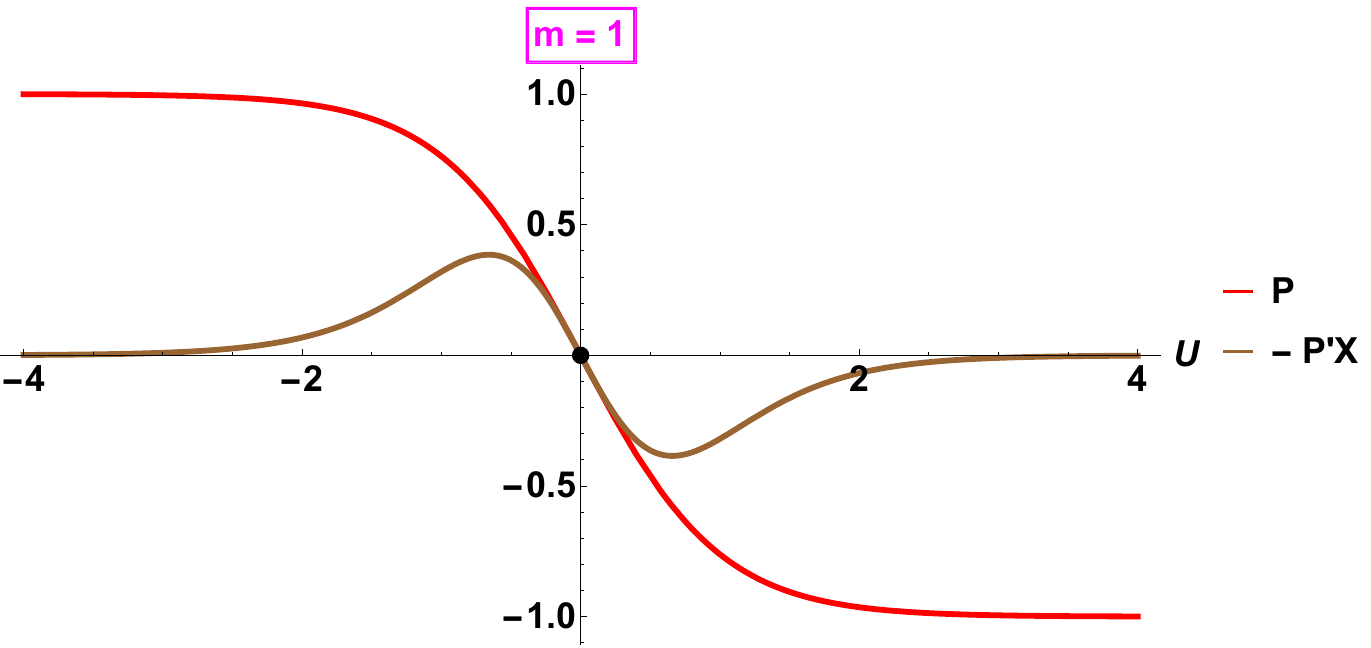}\quad
\includegraphics[scale=.3]{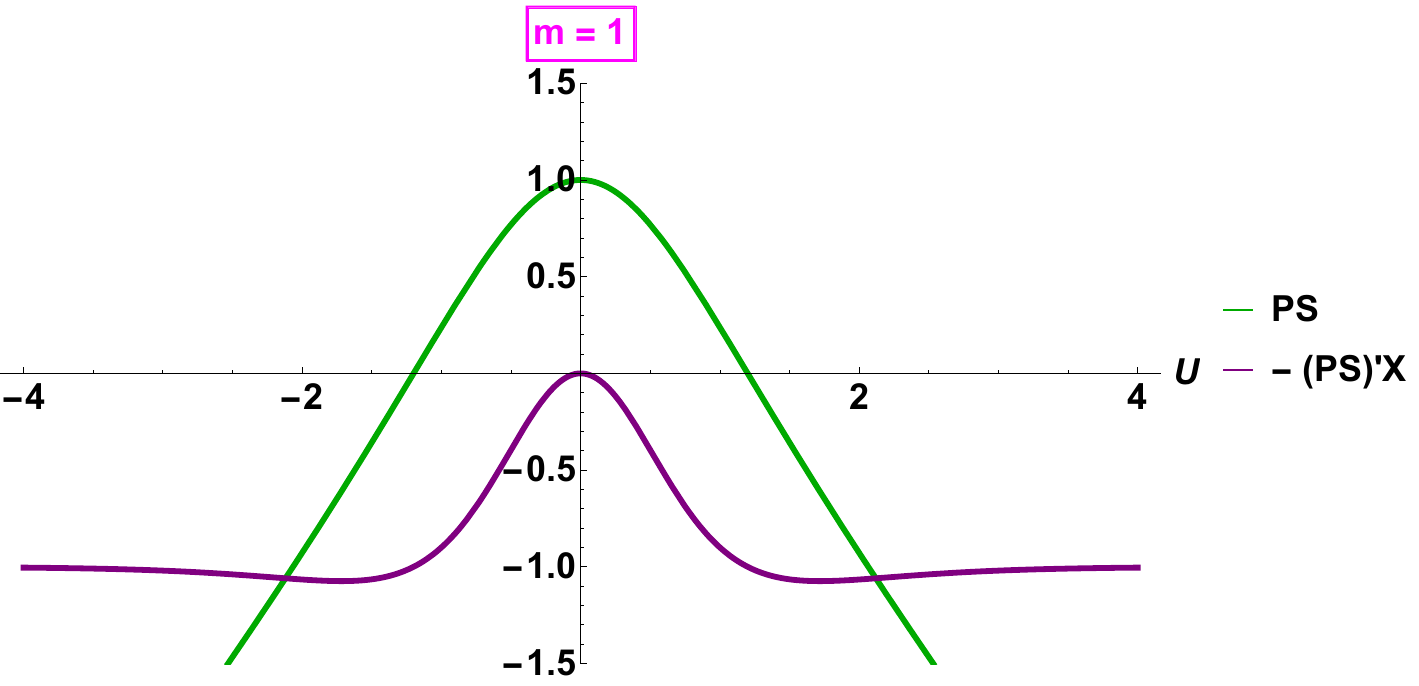}
\vskip-2mm
\hskip-15mm(a) \hskip76mm (b)
\vskip-3mm
\caption{\textit{\small For wave number ${\bf m=1}$, the  
(a) translation generator \red{${\bf P}_1$} [which generates also a trajectory] is {odd}. It has one zero, at $U=0$. (b) the generator of a Carroll boost, \dgreen{${\bf Q}_1 = {\bf P_1S_1}$}, is even and has two, symmetrically positioned zeros. 
}
\label{trans+boost-comp-m1}
}
\end{figure}

\begin{figure}[h]
\includegraphics[scale=.33]{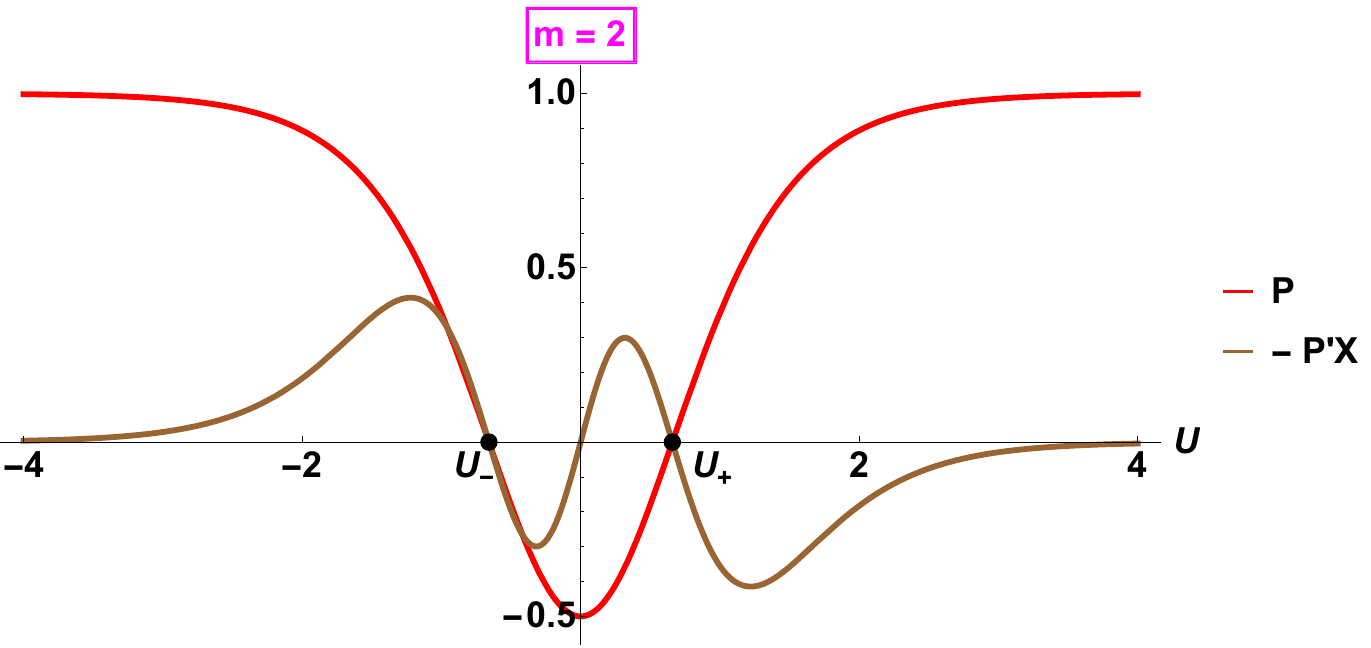}\quad
\includegraphics[scale=.33]{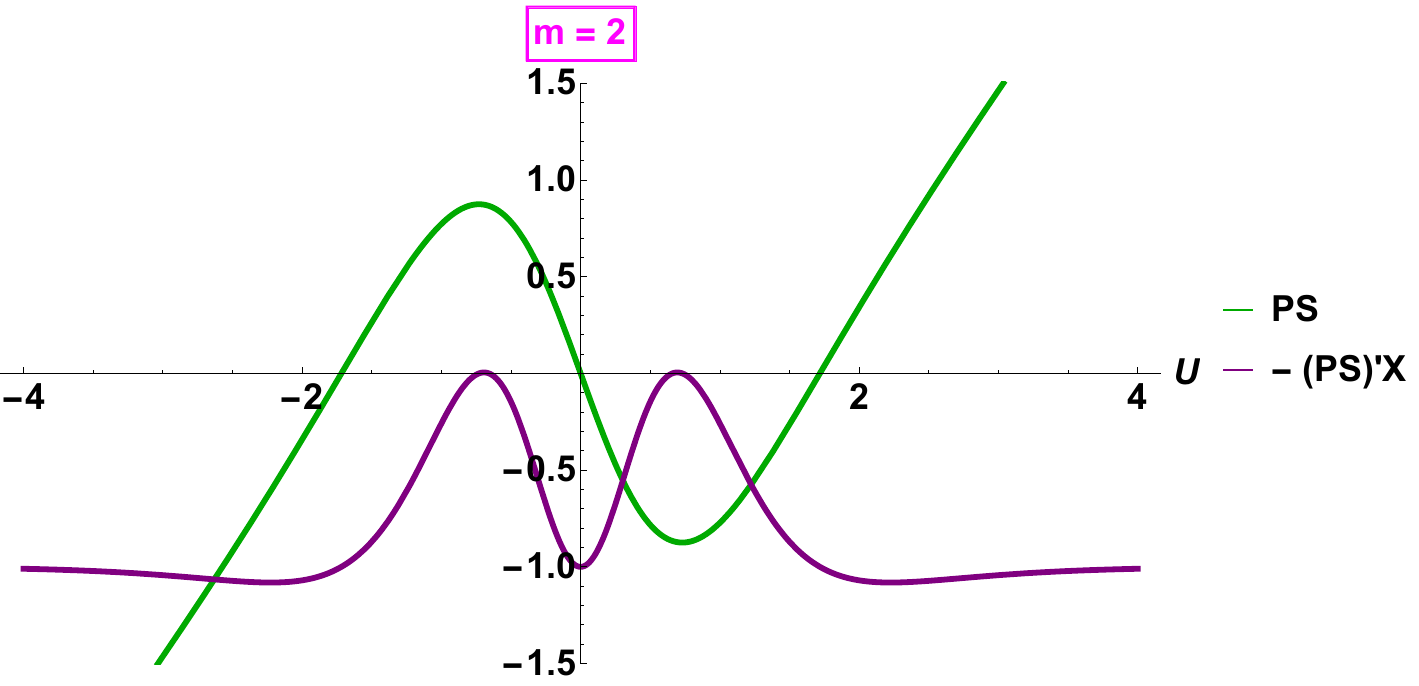}
\vskip-2mm
\hskip-14.5mm(a) \hskip75.5mm (b)
\vskip-4mm
\caption{\textit{\small Infinitesimal Carroll generators \eqref{CarrollPTm2} 
with wave number ${\bf m=2}$.
(a) For a translation, 
the transverse component \red{${\bf P}_2$} is even, and has two zeros, at $U_{\mp}$. (b) For a boost, \dgreen{${\bf Q}_2 = {\bf P_2S_2}$} is odd and has $3$ zeros.  
}
\label{trans+boost-comp-m2} 
}
\end{figure}

\medskip
Turning to symmetries,
the coefficients of the globally defined  Carroll generators $P$ and $Q$ 
in \eqref{CarrollBrink}, listed in 
\eqref{CarrollPTm1} and \eqref{CarrollPTm2} are depicted in FIGs. \ref{trans+boost-comp-m1} and \ref{trans+boost-comp-m2}.
\begin{eqnarray}
\Theta_{m=1}^{Brink}  &=& h \frac{\partial}{\partial V} + c \Big({\tanh U}\,\frac{\partial}{\partial X} -(\sech^2 U )X
\frac{\partial}{\partial V}\Big) 
\nonumber
\\[4pt]
&+&  b\Big({\big(U\tanh U - 1 \big)}\frac{\partial}{\partial X} - \big(\tanh U + U\sech^2 U\big) X \frac{\partial}{\partial V}\Big)\,. 
\label{CarrollPTm1}
\end{eqnarray}
\begin{eqnarray}
 \Theta_{m=2}^{Brink}\!&=&\!h\frac{\partial}{\partial V} + c \left(\big(3\tanh^2U-1\big)\frac{\partial}{\partial X} - \big(6\tanh U\sech^2(U)\big) X \frac{\partial}{\partial V}\right)
\label{CarrollPTm2}
\\[4pt]
&&\hskip-10mm{+}b\left(\frac{1}{4}\big(2U - 3\tanh U - 3U\sech^2U\big) \frac{\partial}{\partial X} 
 - \frac{1}{2}\big(1-3\sech^2 U + 3U\sech^2 U \tanh U\big)X \frac{\partial}{\partial V}\right). \qquad
\nonumber
\end{eqnarray}

%

\section{Graphic representation of the symmetry generators}\label{PlanarSymm}

\begin{figure} [h]%
\vskip-6mm
\includegraphics[scale=.33]{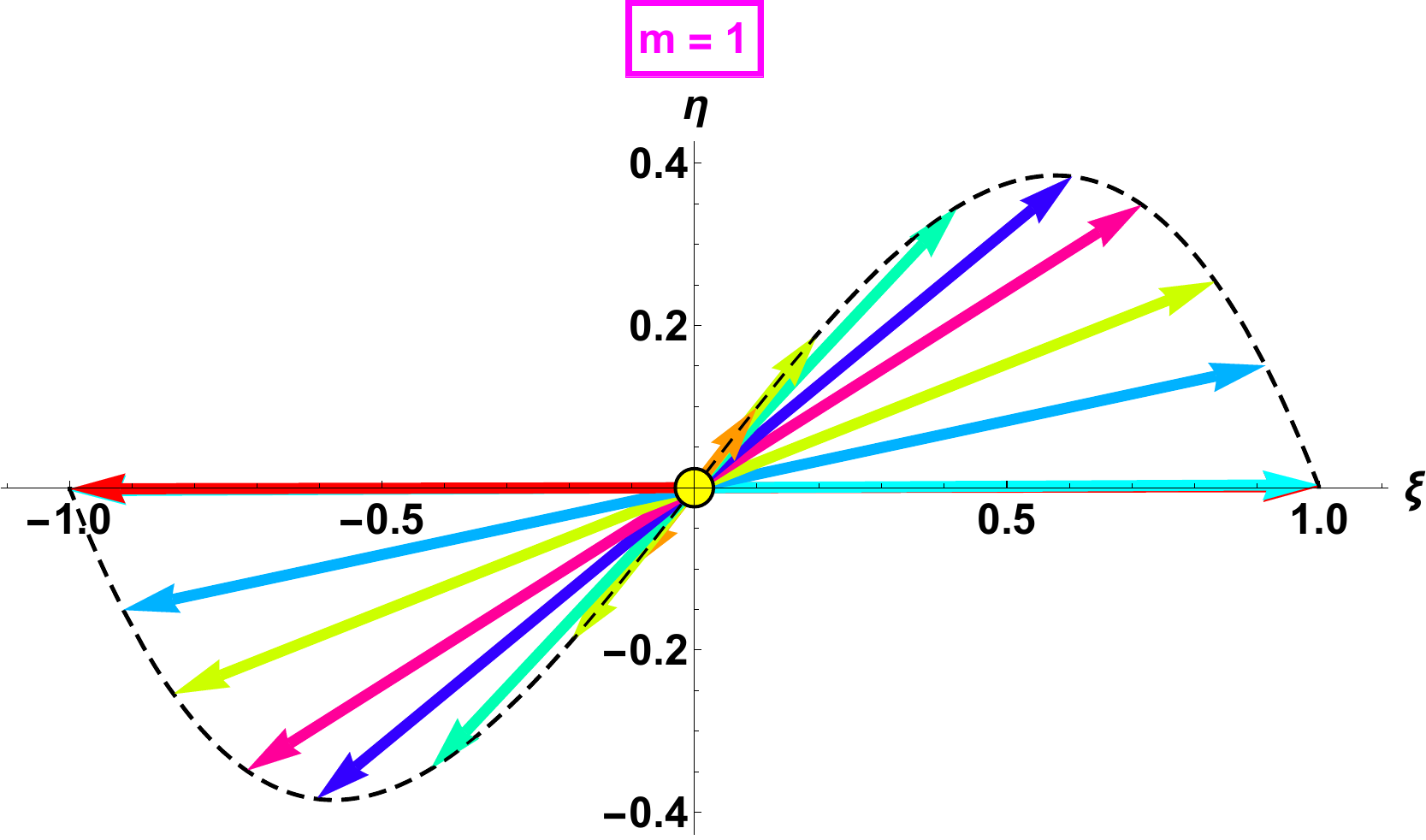}
\vskip-2mm
\caption{\textit{\small Infinitesimal translations \magenta{$(\bxi,\eta)$} along a  DM Brinkmann trajectory  $\bX(U)$ with  ${\bf m=1}$, shown in the co-moving tangent space.
The trajectory is ``hidden'' in the {\bf yellow blob}.
}
\label{2DTransm1}
}
\end{figure}

\begin{figure} [ht]
\includegraphics[scale=.28]{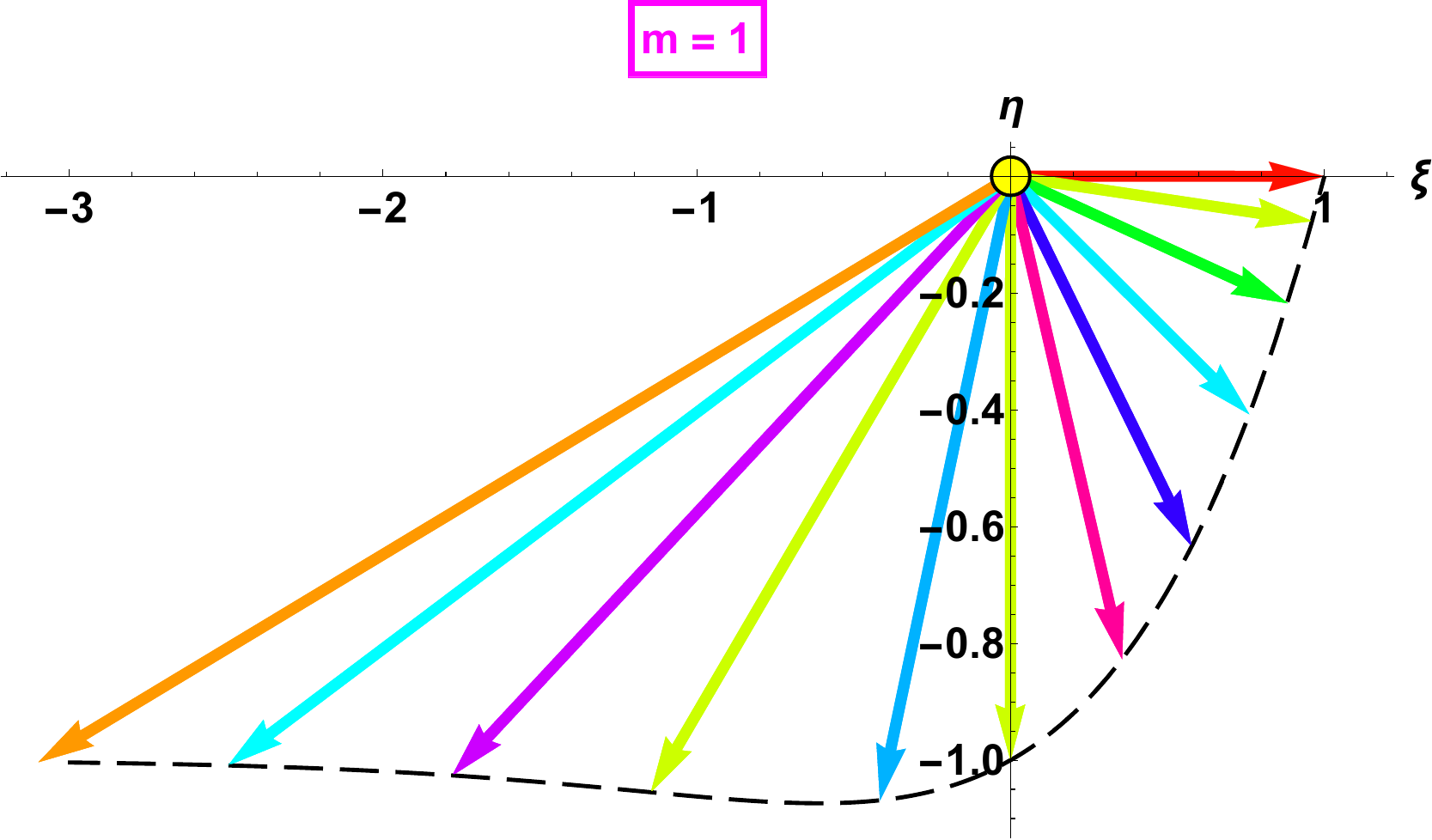} 
\vskip-3mm
\caption{\textit{\small  
Infinitesimal boosts along along the trajectory  ``hidden behind the yellow blob''
with ${\bf m=1}$,  shown in the co-moving tangent space. 
} 
\label{2DBoostm1}
}
\end{figure}

\begin{figure} [h] 
\includegraphics[scale=.28]{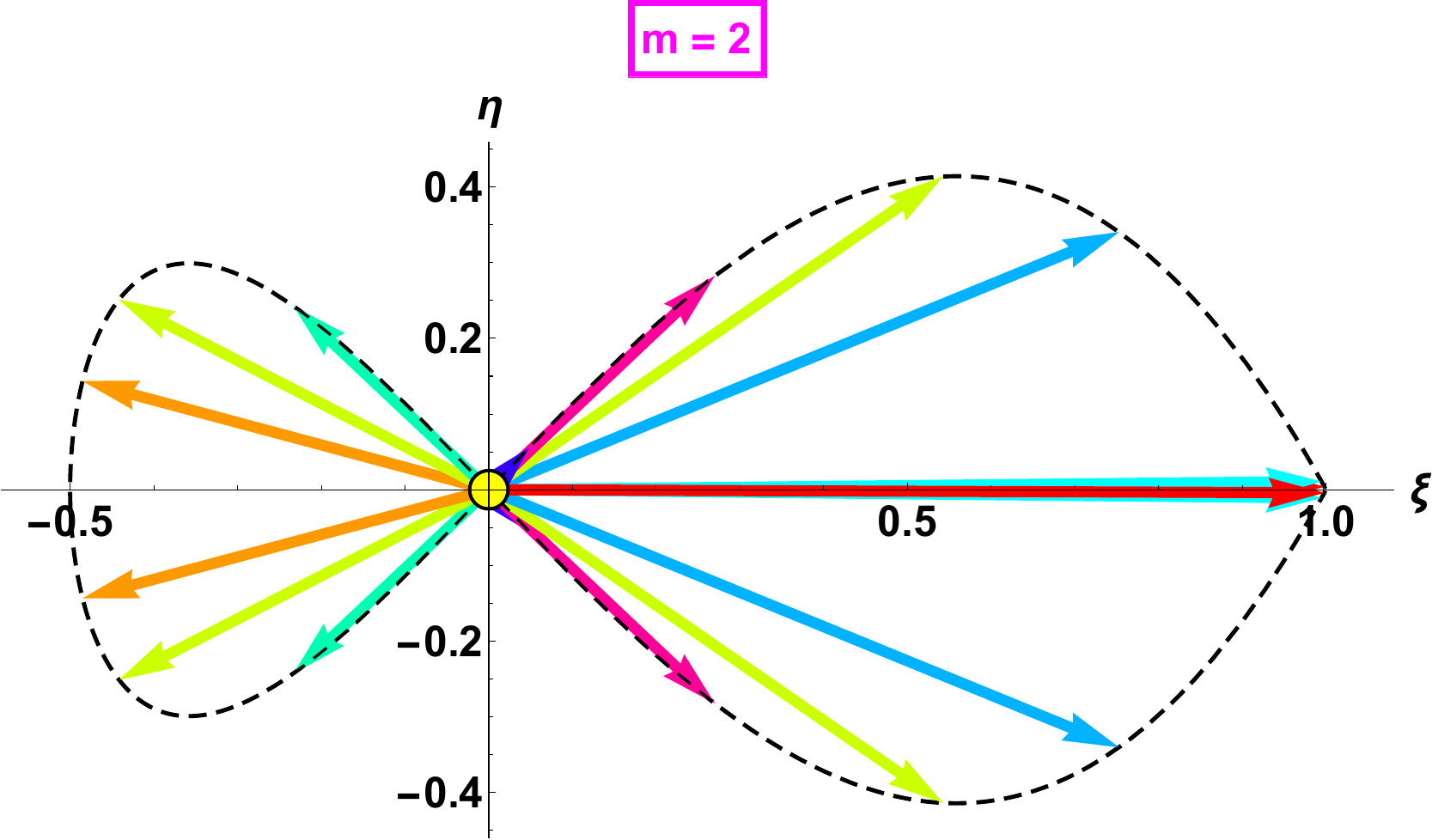}
\vskip-2mm
\caption{\textit{\small 
Translations act along the ${\bf m=2}$ DM trajectory, see  \eqref{CarrollPTm2}.
The  trajectory $\bX(U)$ with  ${\bf m=2}$ is ``hidden" in the yellow blob. 
 } 
\label{2DTransm2}
}
\end{figure}

\begin{figure} [h] 
\includegraphics[scale=.33]{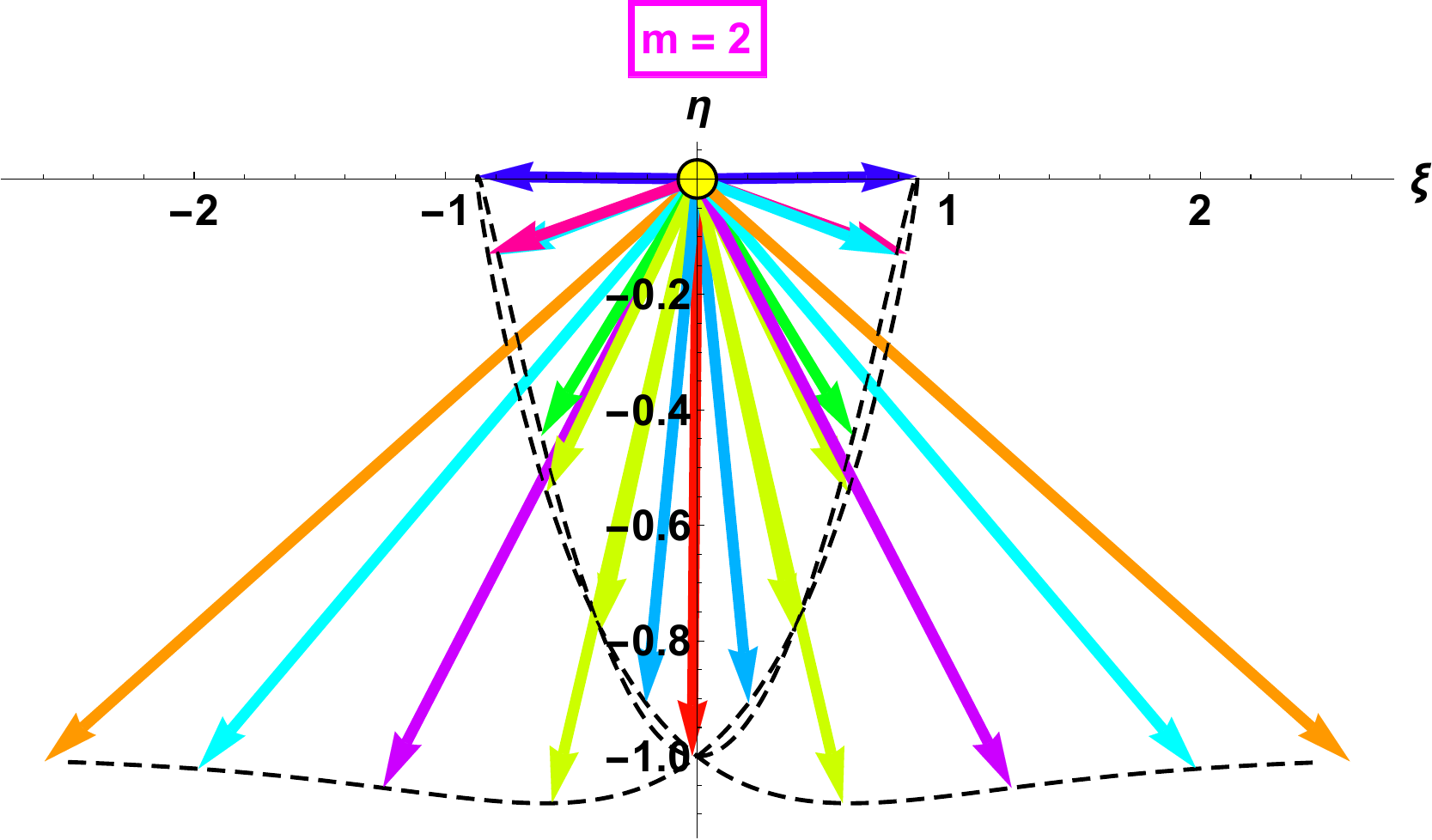}
\vskip-3mm
\caption{\textit{\small Infinitesimal Carroll boosts in Brinkmann coordinates  
along a chosen DM trajectory marked by a yellow blob, with wave number ${\bf m=2}$.   
}
\label{2DBoostm2}
}
\end{figure}

The symmetry-generating vectors along the trajectories
shown in FIGs.\ref{trans+boost-comp-m1} and  \ref{trans+boost-comp-m2} are conveniently viewed in a co-moving tangent space -- which is a plane carried
along the trajectory. In
 FIGs.  \ref{2DTransm1}--\ref{2DBoostm2} the trajectory is  ``hidden'' in \emph{yellow ``blobs''} and the coordinates $\xi$ and $\eta$ represent the coefficients  of the symmetry generators  $\Theta^{Brink}=\bxi\p_X+\eta\p_V$ in 
\eqref{CarrollBrink}, spelt out in \eqref{CarrollPTm1}-\eqref{CarrollPTm2}. 

FIG.\ref{2DTransm1} shows that the translation vector starts at ${\bf U=-\infty}$ from $\brown{\bxi=-1, \eta=0}$. 
It leaves the caustic point for ${\bf U=0}$ at 
$\bX=0,\,V=0$ invariant.
Arriving into the Afterzone, the translation vectors change sign and end, for $U = +\infty$, at $\cyan{\bxi=+1, {\eta}=0}$. Thus they act again as usual translations -- but with reversed sign.

In FIG.\ref{2DBoostm1}, the longest arrow (in \brown{\bf brown}) is both the initial \emph{and} the final boost vector for ${\bf U = \mp\infty}$. The shortest \red{\bf red} arrow \red{$(\bxi=1,{\eta}=0)$} is reached for ${\bf U = 0}$ at the caustic point $\bX=0, V=0$, where the boost shifts all trajectories by the same amount.

According to FIG.\ref{2DTransm2},  
the translation generator (with $c=1/2$) is, for $U=-\infty$, $(\bxi=1,\, \eta=0)$. FIGs. \ref{trans+boost-comp-m1} and \ref{trans+boost-comp-m2} confirm that, consistently with \eqref{CarrollPTm1}, the two focal points of the trajectory, at  $U_{\mp}$, are left invariant.
For $U=+\infty$ the ``8-shaped" curve returns  to 
 $(\bxi=1,\, \eta=0)$, where it started from.

\goodbreak

\begin{figure}[h]
\includegraphics[scale=.4]{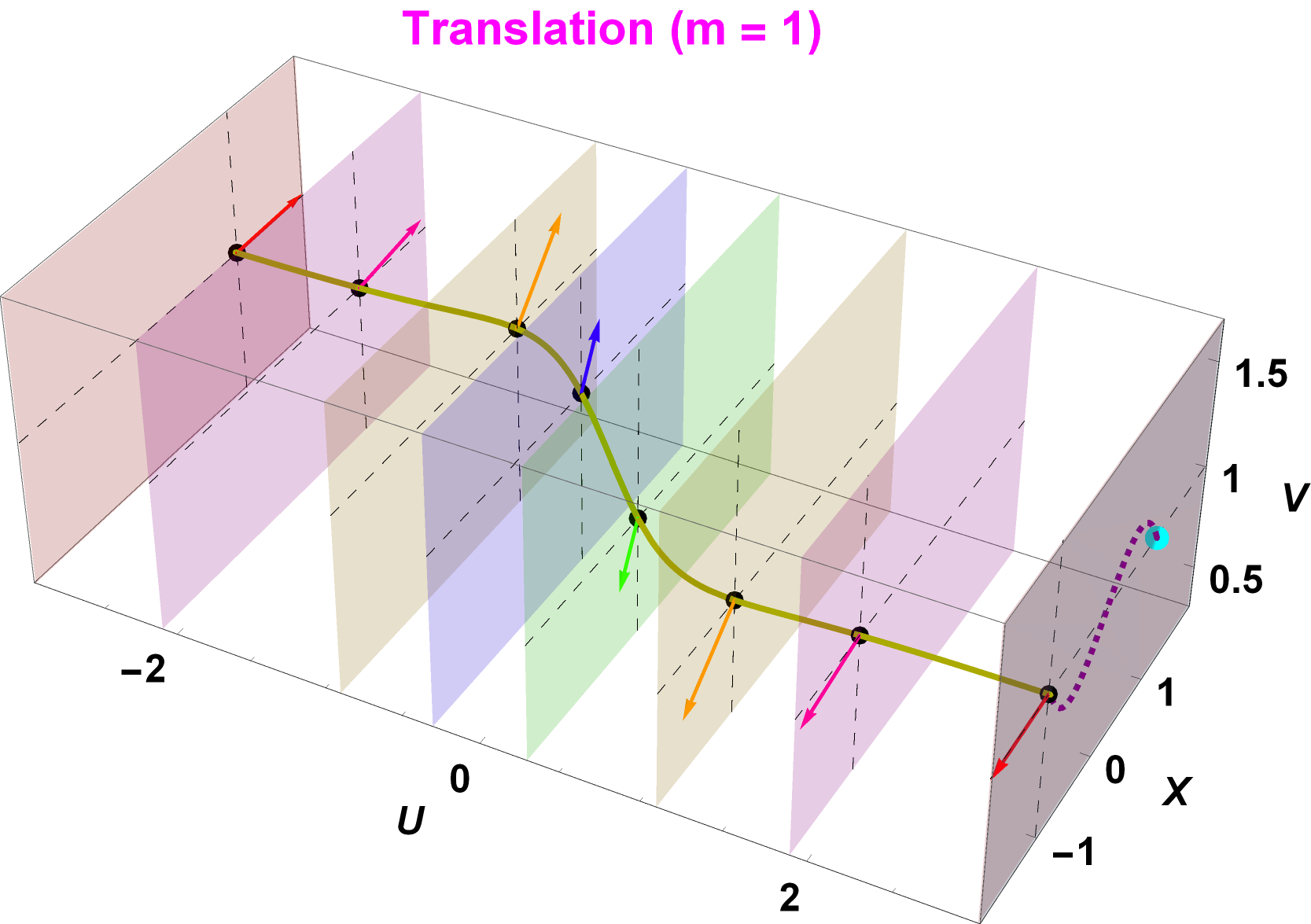}
\\[14pt]
\includegraphics[scale=.4]{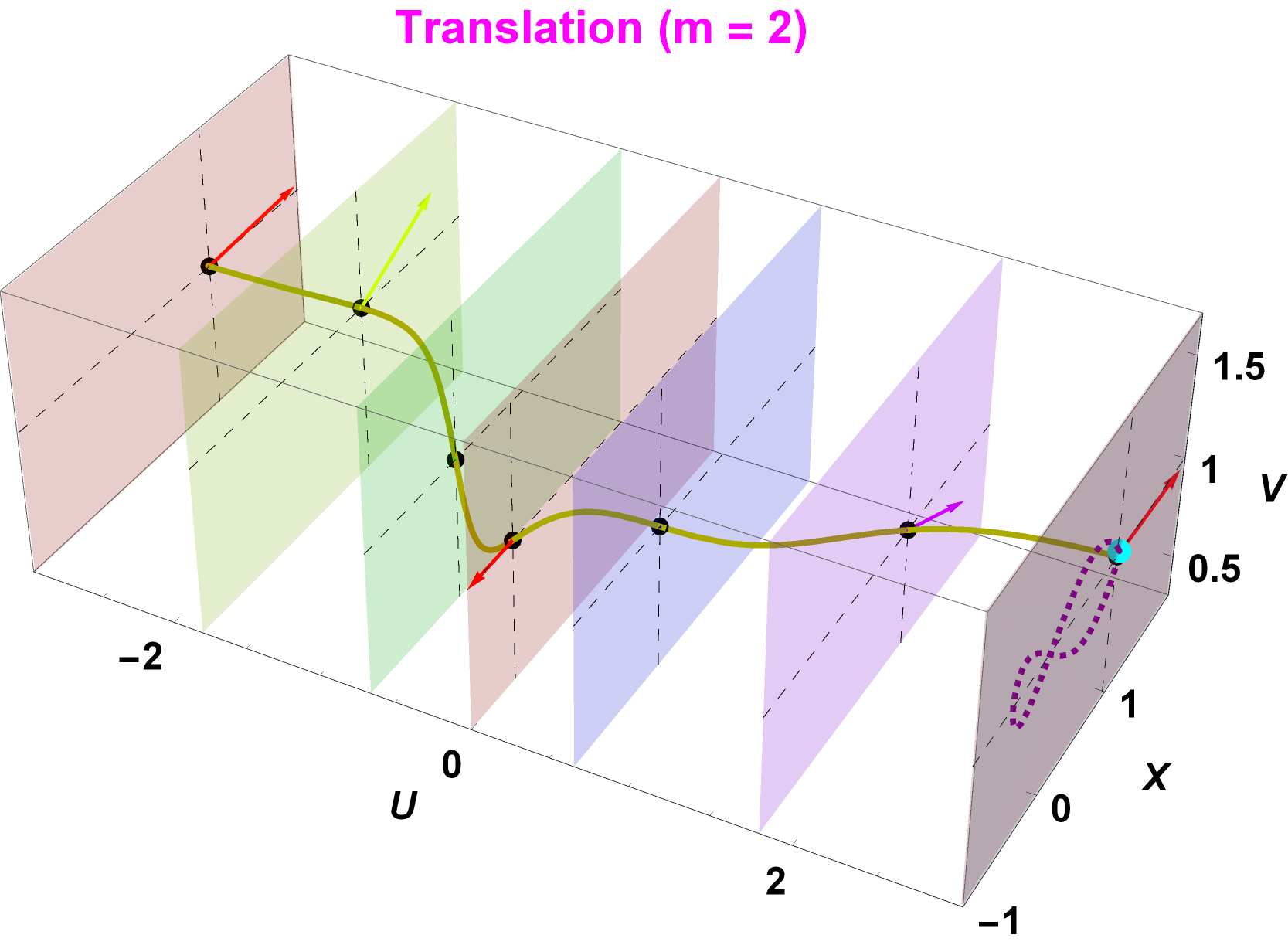}
\vskip-4mm\caption{\textit{\small Translations implemented along the trajectories with wave numbers $m=1$ and $m=2$. 
}
\label{translation3d-m1m2} }
\end{figure}

Boosts along the $m=2$ trajectory are shown in FIG.\ref{2DBoostm2}. The longest vector on the left (in \brown{\bf brown}) is, consistently with \eqref{CarrollPTm1} and \eqref{CarrollPTm2} with $b=1/2$ and $c=2$, the  boost for $U = - \infty$, and the longest one on the right (also in \brown{\bf brown}) is for $U = \infty$. The two shortest \blue{\bf blue} arrows \blue{($\bxi=\pm 1, {\eta}=0$)} show the boost acting as a translations by $\mp b$ at the caustic points $U_\mp$. 
Combining FIGs.\ref{trans+boost-comp-m1} with FIG.\ref{2DTransm1} provides us with the outfolding to 3 dimensions, shown in  FIG.\ref{translation3d-m1m2}.

The translations implemented along the trajectories vanish where \red{$P(U)=0$}. They return to their initial value  for $m$ odd and their transverse component changes sign for $m$ even.

\begin{figure}[h]
\includegraphics[scale=.36]{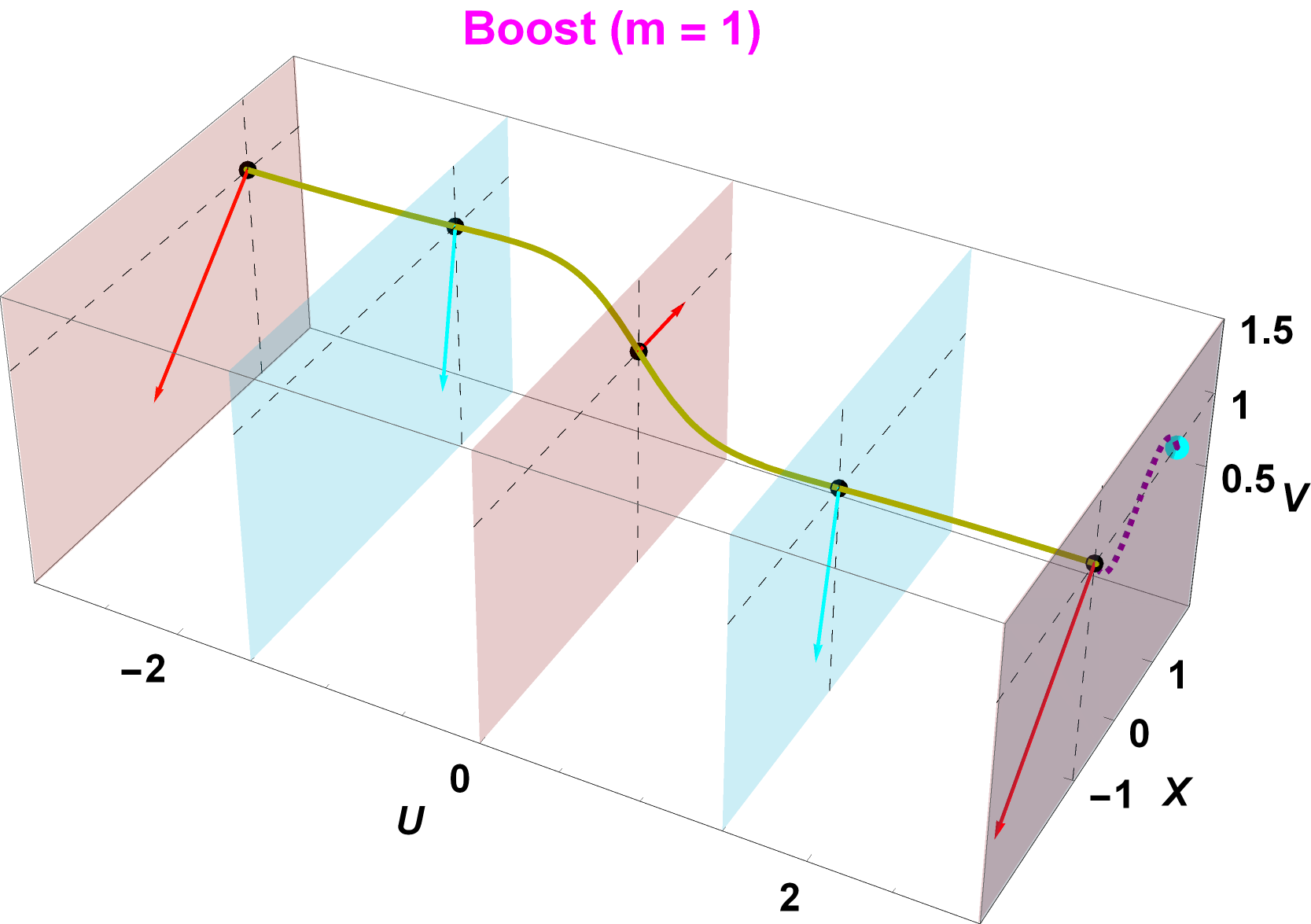}
\\[12pt]
\includegraphics[scale=.38]{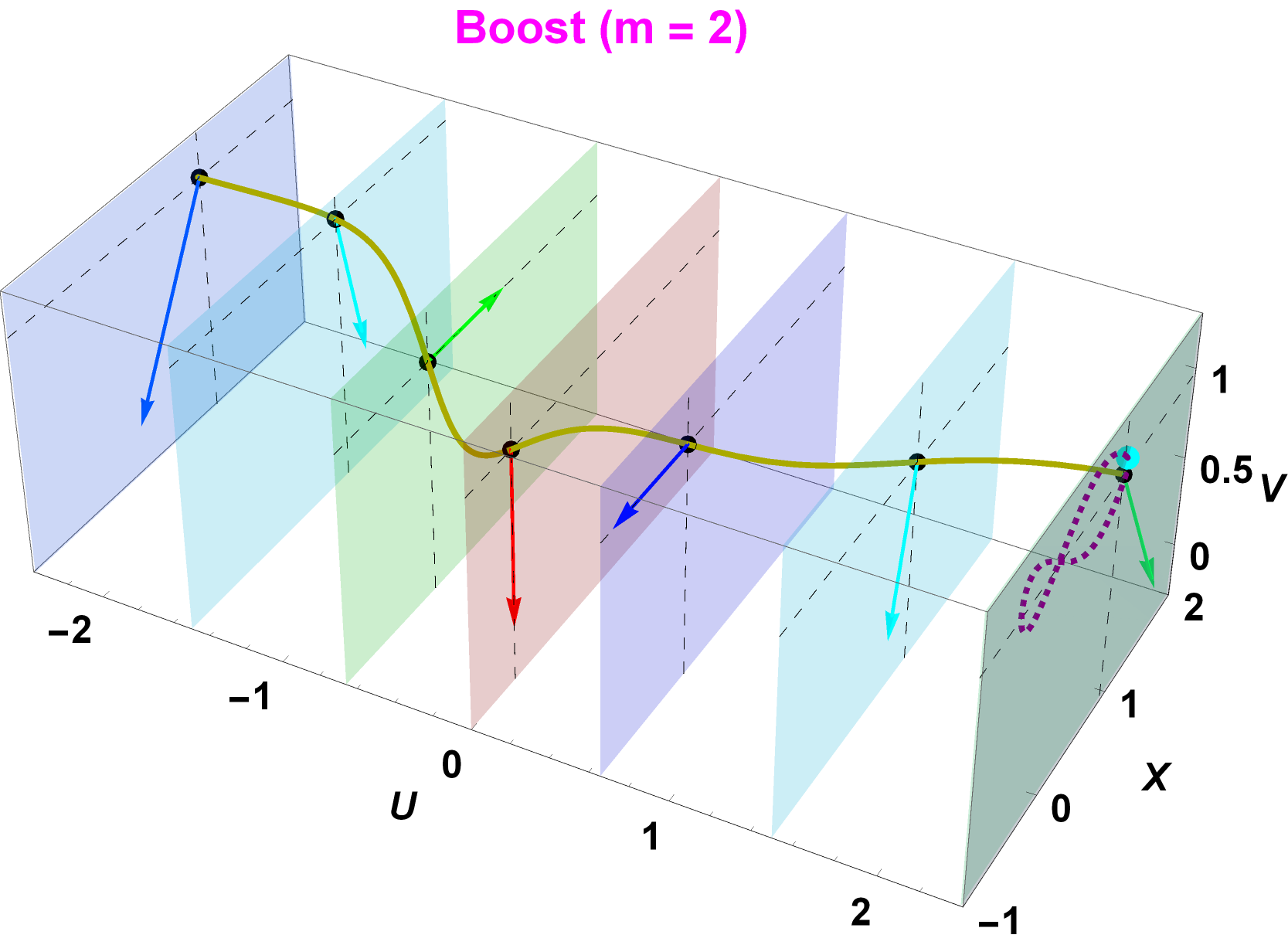}
\vskip-4mm\caption{\textit{\small 
Carroll boosts implemented along the trajectories with wave number 
$m=1$ and $m=2$.
}
\label{boost3d-m1m2} }
\end{figure}
Carroll boosts implemented along the trajectories
 behave as pure translations at \red{$P(U)~=~0$}.
They return to their initial value for $m$ odd and
their transverse component changes sign for $m$ even. 

Conversely, we duly recover the \PT profile \eqref{PTPot}  from the \SL solutions by using the Schwarzian formulas in sect.\ref{SchwarzianSec}. 
 We show this here for the simplest, $m=1$ solution \eqref{BPQm1m2}, with the cast
\begin{equation}
\varphi_1(U) = P(U) = -\tanh U,  \qquad  \varphi_2(U) = Q(U) = 1-U\tanh U\,. 
\label{m1varphi}
\end{equation}
The Wronskian equals to $(-1)$ everywhere. 
Then inserting $f'(U)=-\cosh^2 U+1\,$
into the Schwarzian
 \eqref{Schwarzprofile}  reproduces the $m=1$ profile  $\cA(U) = - \frac{8}{\cosh^2(U)}$ in \eqref{PTPot}.
\goodbreak

\section{Conclusion}\label{Conc}

Our results shed further light on  the intimate relation of Carroll symmetry and the Memory Effect~: the  matrix $P$ with initial conditions \eqref{Pinit} yields  our geodesics. The general expression \eqref{XPQ} which involves both DM and VM is one of our main results.

When the parameter $m$ is an integer, then the wavezone accommodates $m$ half-waves and we get DM. For $m=2\ell+1$
$P(U)$ is odd and we do get net displacement. For  $m=2\ell$, though, $P(U)$ is even and 
the particle returns, after non-trivial motion in the wave zone, to its initial position~: we get DM with no final displacement, consistently with \eqref{BDMtraj} and confirmed by FIGs.\ref{Btrajm1m2}, \ref{trans+boost-comp-m1}a and \ref{trans+boost-comp-m2}a. 
The  Brinkmann trajectories are focused at the zeros of $\det(P)$.

DM trajectories  have \emph{zero momentum} and \emph{zero non-relativistic energy} \cite{DM-1}. They can be found by putting  $\bp_0=0$ into  \eqref{XPQ}. 
Turning on the momentum, $\bp_0\neq0$, converts the trajectory to VM.
\goodbreak

\medskip
\goodbreak
Both BJR and Brinkmann coordinates have their advantages and drawbacks.

The merit of BJR is simplicity \cite{Sou73,Carroll4GW}.
Our initial conditions \eqref{Pinit} imply $\bp=0$,  
and the ``motion'' reduces merely to a fixed point determined by the conserved boost momentum $\bk=\bx_0$ and $v_0$ ---  the initial position  \eqref{BJRtraj}-\eqref{BJRnomotion}. The dynamics is carried by $P(u)$.
The price to pay is that the BJR description is valid only between the {mandatory}  zeros of $\det(P)$, where the $S$-matrix diverges \cite{Sou73,LongMemory,EZHRev}, and solutions in the adjacent domains have to be fitted together \cite{EZHRev}. 

The advantage of the Brinkmann form \eqref{ABXeq}--\eqref{ABVeq} is that it replaces the  singular Souriau matrix $S$  in \eqref{Smatrix} by the globally defined  \StL solutions $P$ and $Q=PS$,  \eqref{SL+cond} and \eqref{2ndSLsol}, respectively.
For zero momentum, \eqref{0mom}, the  non-DM solution $Q$ is turned off, leaving us with $P$ alone which plays its second, trajectory-role.
 The drawback  is that finding those   solutions may  be  laborious \cite{QLdPT,DM-1,Sila-PLB}.
 In conclusion, the main difficulty boils down, in \emph{both} coordinate systems, to solving a \StL equation.
\goodbreak

Switching to Brinkmann coordinates allowed us to extend the locally given Carrollian Killing vectors \cite{Sou73} to the whole spacetime. These vectors are directly related to geodesics and allow us to unify the Displacement and Velocity Memory effects by combining the two independent solutions $P$ and $Q$ of the \SL problem, see  \eqref{XPQ}, highlighting the intimate relationship of geodesics with Carroll symmetry  \cite{EZHRev,GenRelGrav}.

The Brinkmann and BJR forms have also been compared by
Harte \cite{Harte} from a different point of view. Showing that the BJR plane wave metric is a generalisation of perturbative weak gravitational waves, he discussed memory effects in a perturbative manner.     
  
Figs. \ref{2DTransm1} -- \ref{boost3d-m1m2} illustrate the curious behavior of the Carroll generators along the trajectories.

Further insight is provided by the Schwarzian framework  \cite{Ovsienko,ZZH,Vlahakis}. This approach confirms that $Q=PS$ is indeed a globally defined, 2nd solution of the \SL problem which does not involve the  singular Souriau matrix $S$. 

Our results go beyond the toy dimension $D=1$ dimensional profiles and the \PT example. Similar results hold for the Scarf profile in $D=2$, which provides us with an analytic approximation of the physically relevant flyby case \cite{ZelPol,Sila-PLB}. 

Gravitational waves and their observable effects are also formulated by means of Bondi metric \cite{Winicour} 
which relates the gravitational emission and the mass loss. The Bondi metric is axially symmetric and asymptotically flat. In general, our plane waves are not asymptotically flat, though. 
  However, both metrics share similar properties like having a covariantly constant null Killing vector which determines the direction of wave propagation.
The transverse part of the Bondi metric is quadratic in the radial coordinate which defines the luminosity distance. In \cite{Harte}, the BJR plane wave metric is used to compute similar physical properties.    
To clarify the differences and the similarities between the Bondi formalism with ours here requires more effort and will be reported elsewhere. 

Supersymmetry \cite{MSUSY} sheds further light on the Memory Effect.
Further details will be presented in a comprehensive review~\cite{ZEHPR}.
\goodbreak

\kikezd{Note added in proof}.
After having submitted our paper, we realized  that our decomposition \eqref{XPQ} spelt out for \PT can actually be generalized. It applies in particular  also to the \dPT profile \cite{DM-1,DM-2} with the cast: $P$ is the \DM solution of the Sturm-Liouville equation \eqref{ABXeq} solved, for dPT, in terms of confluent Heun functions. The second independent solution  $Q = PS$, involves also the Souriau matrix $S$ \eqref{Smatrix} \cite{ZEHPR,QL-MdL}.

\acknowledgements{
The authors are indebted to J. Balog, J. Ben Achour and Z. Silagadze for correspondance.
PMZ was partially supported by the National Natural Science Foundation of China (Grant No. 11975320).
ME is supported by Istanbul Bilgi University research fund (BAP) with grant no: 2024.01.009 and by The Scientific and Technological Research Council of Turkey (T\"{U}BITAK) under grant number 125F021. PMZ and PAH benefitted of T\"{U}BITAK 2221 grant and hospitality at Bilgi University during their visit in Istanbul.
}
\goodbreak


\end{document}